\documentclass[%
superscriptaddress,
twocolumn,
 amsmath,amssymb,
 aps,
prb,
]{revtex4-1}

\usepackage{graphicx}
\usepackage{dcolumn}
\usepackage{bm}
\usepackage{natbib}

\usepackage{amsmath} 

\bibpunct{[}{]}{,}{n}{}{}
\begin{document}

\preprint{APS/123-QED}

\title{Optical excitation of propagating magnetostatic waves\\in an epitaxial Galfenol film by an ultrafast magnetic anisotropy change}

\author{N. E. Khokhlov}
\email{n.e.khokhlov@mail.ioffe.ru}
 \affiliation{Ioffe Institute, 194021 St. Petersburg, Russia}
\author{P. I. Gerevenkov}
 \affiliation{Ioffe Institute, 194021 St. Petersburg, Russia}
 \author{L. A. Shelukhin}
 \affiliation{Ioffe Institute, 194021 St. Petersburg, Russia}
 \author{A. V. Azovtsev}
 \affiliation{Ioffe Institute, 194021 St. Petersburg, Russia}
  \author{N.~A.~Pertsev}
 \affiliation{Ioffe Institute, 194021 St. Petersburg, Russia}
  \author{M. Wang}
 \affiliation{School of Physics and Astronomy, The University of Nottingham, NG7 2RD Nottingham, UK}
 \author{A. W. Rushforth}
 \affiliation{School of Physics and Astronomy, The University of Nottingham, NG7 2RD Nottingham, UK}
\author{A. V. Scherbakov}
 \affiliation{Ioffe Institute, 194021 St. Petersburg, Russia}
 \affiliation{Experimental Physics II, Technical University Dortmund, D44227, Dortmund, Germany}
\author{A. M. Kalashnikova}
 \affiliation{Ioffe Institute, 194021 St. Petersburg, Russia}

\date{\today}

\begin{abstract}
Using a time-resolved optically-pumped scanning optical microscopy technique we demonstrate the laser-driven excitation and propagation of spin waves in a 20-nm film of a ferromagnetic metallic alloy Galfenol epitaxially grown on a GaAs substrate.
In contrast to previous all-optical studies of spin waves we employ laser-induced thermal changes of magnetocrystalline anisotropy as an excitation mechanism. 
A tightly focused 70-fs laser pulse excites packets of magnetostatic surface waves with an $e^{-1}$-propagation length of 3.4 $\mu$m, which is comparable with that of permalloy. As a result, laser-driven magnetostatic spin waves are clearly detectable at distances in excess of 10 $\mu$m, which promotes epitaxial Galfenol films to the limited family of materials suitable for magnonic devices.
A pronounced in-plane magnetocrystalline anisotropy of the Galfenol film offers an additional degree of freedom for manipulating the spin waves' parameters.
Reorientation of an in-plane external magnetic field relative to the crystallographic axes of the sample tunes the frequency, amplitude and propagation length of the excited waves.


\end{abstract}

\maketitle



\section{\label{sec:Intro}Introduction}

In magnonics coherent spin waves (SWs) are employed for encoding, transferring, and processing information \cite{Lenk-PhysRep2011,Nikitov:UFN2015,ChumakJPD2017}.
The use of SWs enables scaling of magnonic elements down to the nanometer range owing to short wavelengths and the reduction of Joule heating associated with the charge transfer in conventional electronics.
Moreover, an extended functionality of magnonic devices is provided by the possibility to manipulate the amplitudes, phases, and wavevectors of SWs. 
Progress in the field of magnonics relies on the development of approaches to generate and transfer SWs in a controllable manner. 
Efficient conversion mechanisms between electrical/optical pulses and collective magnetic excitations are required to couple magnonic elements to electronic and photonic units \cite{Demidov:2016,Ralph:JMMM2008,Hellman:RMP2017}.
For rapid growth of the field of magnonics, it is essential to extend the range of materials and structures supporting long SWs propagation distances and enabling their control \cite{Grundler:NPhys2015,Davies:LTP2015,ChumakNPhys:2015,Hamalainen:PRL2018,Sadovnikov:PRL2018}.

Optical radiation allows the magnetic parameters of materials to be altered reversibly at various timescales down to femtoseconds \cite{Kovalenko:UFN1986,Kirilyuk:2010}.
This has led to the emergence of a photo-magnonics \cite{Lenk:2013,VogelNPhys:2015,VogelSciRep:2018,Au:2013,Satoh:2012,Bossini:NComm2016,Chang:PRAppl2018}, where laser pulses 
are employed as a tool for both driving SWs and manipulating their propagation. 
In particular, it has been demonstrated that femtosecond laser pulses enable excitation of SWs with controlled wavevectors and propagation directions \cite{Satoh:2012,SavochkinSciRep:2017,JacklPRX:2017,Chernov:2017}.
However, up to now the effects of short laser pulses employed to drive SWs have been limited to ultrafast opto-magnetic phenomena \cite{JacklPRX:2017,Satoh:2012,Chernov:2017,Hashimoto:2017,JacklPRX:2017,SavochkinSciRep:2017,YoshimineStupakiewicz:JAP2014}, ultrafast demagnetization \cite{Au:2013,IihamaPRB:2016,KamimakiPRB:2017,YunAPEX:2015,ChenSciRep2017}, and coherent energy transfer from elastic waves to the magnon subsystem \cite{Hashimoto:2017,HashimotoPRB:2018, OgawaPNAS:2015, HashimotoAPL:2018, Hashimoto:PhysRevApplied2019}.
These mechanisms place constraints on the properties of the media, the laser pulse parameters, and the excitation geometries, while the excitation of propagating SWs by other ultrafast magnetic phenomena \cite{Kirilyuk:2010} remains unexplored.
Furthermore, the range of materials where the optical generation of SWs has been realized is also very limited and, in fact, coincides with the known suitable media for magnonics \cite{ChumakJPD2017}.

In this Article we examine the feasibility and advantages of excitation of propagating SWs in an anisotropic ferromagnetic film by ultrafast laser-induced thermal changes of the magnetocrystalline anisotropy.
Using time-resolved optically pumped scanning optical microscopy (TROPSOM) \cite{Au:2013}, we reveal propagating  magnetostatic surface waves (MSSWs) excited by a femtosecond laser pulse in a thin film of a ferromagnetic metallic alloy Galfenol (Fe$_{0.81}$Ga$_{0.19}$) epitaxially grown on a GaAs substrate.
We demonstrate that the propagating MSSWs packets are launched via the laser-induced thermal decrease of the magnetic anisotropy occurring on a picosecond time scale and localized within the excitation spot.
Owing to this excitation mechanism, the MSSWs are excited in a simple geometry with an in-plane external magnetic field, and their characteristics can be controlled by the orientation of the in-plane field with respect to the magnetocrystalline anisotropy axes of the film.
We show that the 20-nm thick Galfenol film supports an $e^{-1}$-propagation length of MSSWs as large as 3.4\,$\mu$m, comparable to that of Permalloy -- a model metallic material for magnonics  \cite{ChumakJPD2017}.
Our results promote epitaxial Galfenol to the limited family of materials for magnon-spintronics, reconfigurable magnonics, and ultrafast photo-magnonics.
Furthermore, the ultrafast thermal changes of magnetic anisotropy as a driving mechanism for the excitation of propagating SWs can be applied to a broad range  of materials without specific constraints imposed on their electronic and magnetic structures \cite{BaranovPG:UFN2018}.

The Article is organized as follows. In Sec.\,\ref{sec:Experiment} we describe the FeGa/GaAs sample and the details of the TROPSOM experimental setup. 
In Sec.\,\ref{sec:results} we first present the experimental results on laser-induced excitation of the magnetization precession and SWs (\ref{sec:SWpackets}), discuss the excitation mechanism (\ref{sec:parameters}), the main parameters and features of the MSSWs' propagation (\ref{sec:Lprop}), and reconstruct the MSSWs' dispersion from experimental data (\ref{sec:FFT}).
This is followed by a theoretical analysis of the MSSW dispersion relation specific to the anisotropic FeGa film, calculations of the main MSSWs' parameters and their comparison to those obtained experimentally (\ref{sec:theory}). 
In Sec.\,\ref{sec:Conclusion} we summarize our findings and discuss their possible impact on the field of magnonics.

\section{\label{sec:Experiment}Experimental details}
\subsection{\label{sec:sample}Sample}

For our study, we chose a film of Fe$_{0.81}$Ga$_{0.19}$ with a thickness $d=20$\,nm epitaxially grown on a 350-$\mu$m thick (001)-GaAs substrate by magnetron sputtering, as described elsewhere \cite{Linnik:PhysScr2017}.
The Galfenol film was capped with 3-nm thick Al and 120-nm thick SiO$_2$ protective layers.
The back-side of the substrate was polished to optical quality.
X-ray diffractometry showed that the Galfenol film has a mosaic structure with grain sizes of $\sim$12\,nm, and their crystallographic axes misorientation of $\sim$1.3\,deg.
It is well established that thin films of iron and iron-based alloys epitaxially grown on GaAs exhibit intrinsic cubic and substrate-induced in-plane uniaxial magnetic anisotropies \cite{KrebsJAP:1987,GesterJAP:1996,Wastlbauer:AP2005,HindmarchSpin:2011}.
In particular, in Galfenol films on (001)-GaAs substrates the uniaxial anisotropy axis emerges along the [110] direction \cite{Linnik:PhysScr2017,DanilovPRB:2018}.

\subsection{\label{sec:setup}Experimental setup}

\begin{figure} 
    \center{\includegraphics[width=1\linewidth]{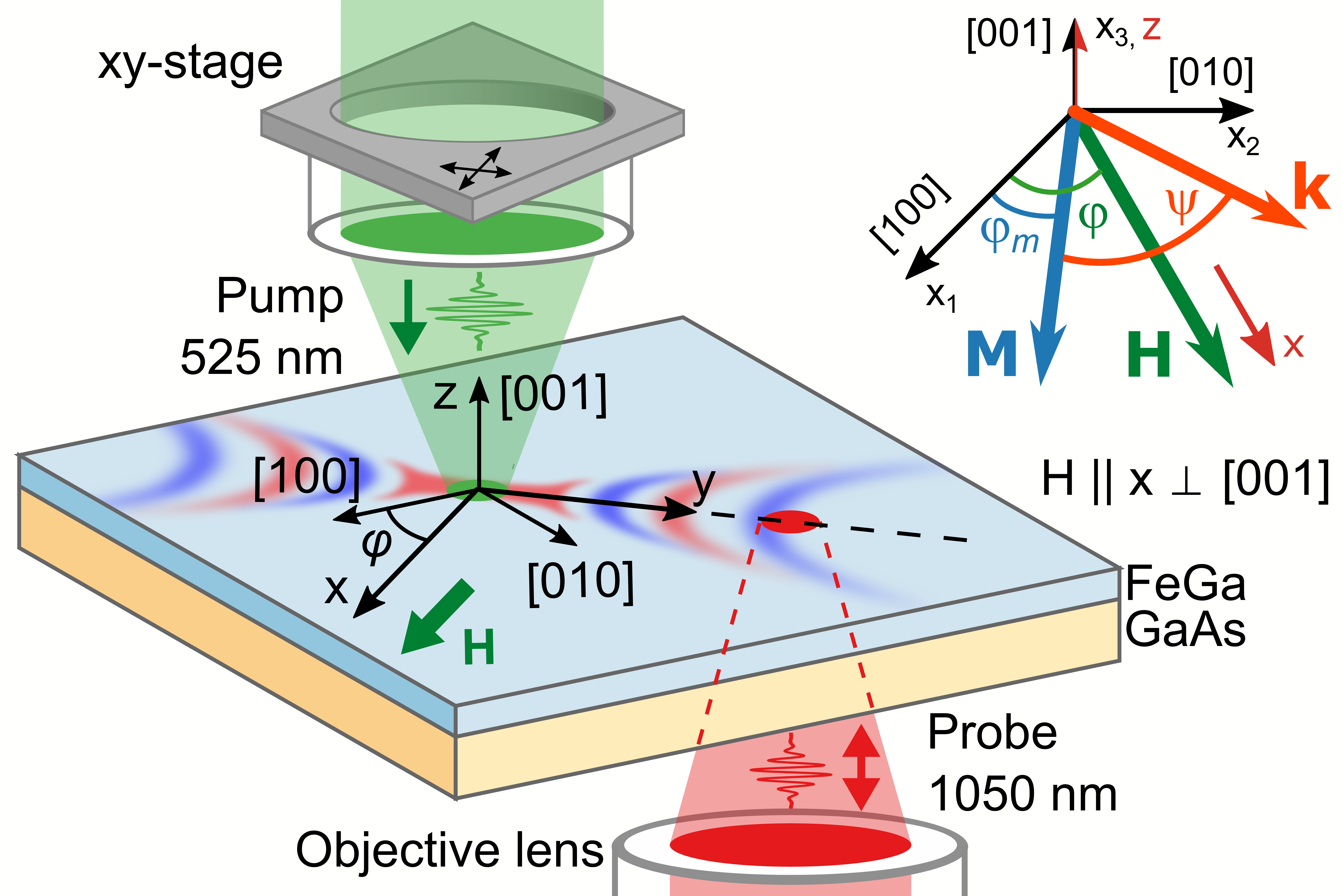}} 
    \caption{\label{fig1}
  Experimental geometry. 
  A DC magnetic field $\mathbf{H}$ is applied along the $x$ axis, while the azimuthal orientation of the sample $\varphi$ can be varied.
  The relative pump-probe distance is changed along either the $x$- or $y$-direction to detect propagation of BVMSWs or MSSWs modes, respectively.
  The blue-red pattern represents SW propagation in the $xy$ plane schematically.
  Inset shows the $x_1x_2x_3$ reference frame linked to the crystallographic axes of the sample, and the angles $\varphi_m,\varphi,\psi$ defining the in-plane orientations of the magnetization $\mathbf{M}$, external DC magnetic field $\mathbf{H}$, and the SW wavevector $\mathbf{k}$, respectively.
  }
\end{figure}

Optically-excited SWs were studied using the TROPSOM setup which enables detection of the spatial-temporal evolution of the polar magneto-optical Kerr rotation $\Delta\theta_k$ proportional to the transient changes of the out-of-plane magnetization component $M_z$ [Fig.\,\ref{fig1}]. 
Here $xyz$ is the laboratory frame with the $z-$axis directed along the sample normal and the $x-$axis chosen to be along the direction of an external DC magnetic field {\bf H}. 
Optical pulses with nominal duration of 70~fs, central wavelength of 1050~nm, and 70~MHz repetition rate generated by the Yb-doped solid-state oscillator laser system were split into pump and probe parts.
The central wavelength of the pump pulses was converted to 525~nm using a $\beta$-BaB$_2$O$_4$ crystal.
The amplitude of the pump pulses was periodically modulated at a frequency of 84\,kHz using a photoelastic modulator 
placed between two crossed Glan-Taylor prisms. 
Two microscope objectives were used to focus the pump and probe pulses onto the Galfenol film from the cap and the substrate sides, respectively.
The pump and probe radii $\sigma$, i.e. the half-width of the beam's spatial profile at which the intensity drops by $\sqrt{e}$, were measured to be $0.8\,\mu$m using the knife-edge method.
The pump fluence was 3.5\,mJ/cm$^2$, and the probe fluence was approximately 20 times lower.
The microobjective for the probe pulses was fixed.
The microobjective for the pump pulses was mounted on the piezoelectric stage, which moves in the $xy$ plane and controls the relative spatial displacements $\Delta x,\,\Delta y$ between the pump and probe pulses.
The pump-probe temporal delay $t$ was controlled by a delay line in the pump pulse optical path.
In the experiments the temporal dependences $\Delta\theta_k(t)$ of probe were obtained at various pump-probe displacements $\Delta x,\Delta y$, and at various azimuthal orientations of the sample defined by the angle $\varphi$ between the $[100]$ crystallographic axis and the $x$-axis.
The probe polarization rotation $\Delta\theta_k(t)$ was detected using a conventional scheme with a Wollaston prism and a balanced photodetector. 
The signal from the photodetector was registered using a lock-in technique with the reference frequency of the pump amplitude modulation.
Thus, only the probe polarization rotation related to the pump-induced dynamics was detected.
The measurements of SWs propagation were performed at room temperature and $H$ = 100 mT.

\begin{figure*} 
    \center{\includegraphics[width=1\linewidth]{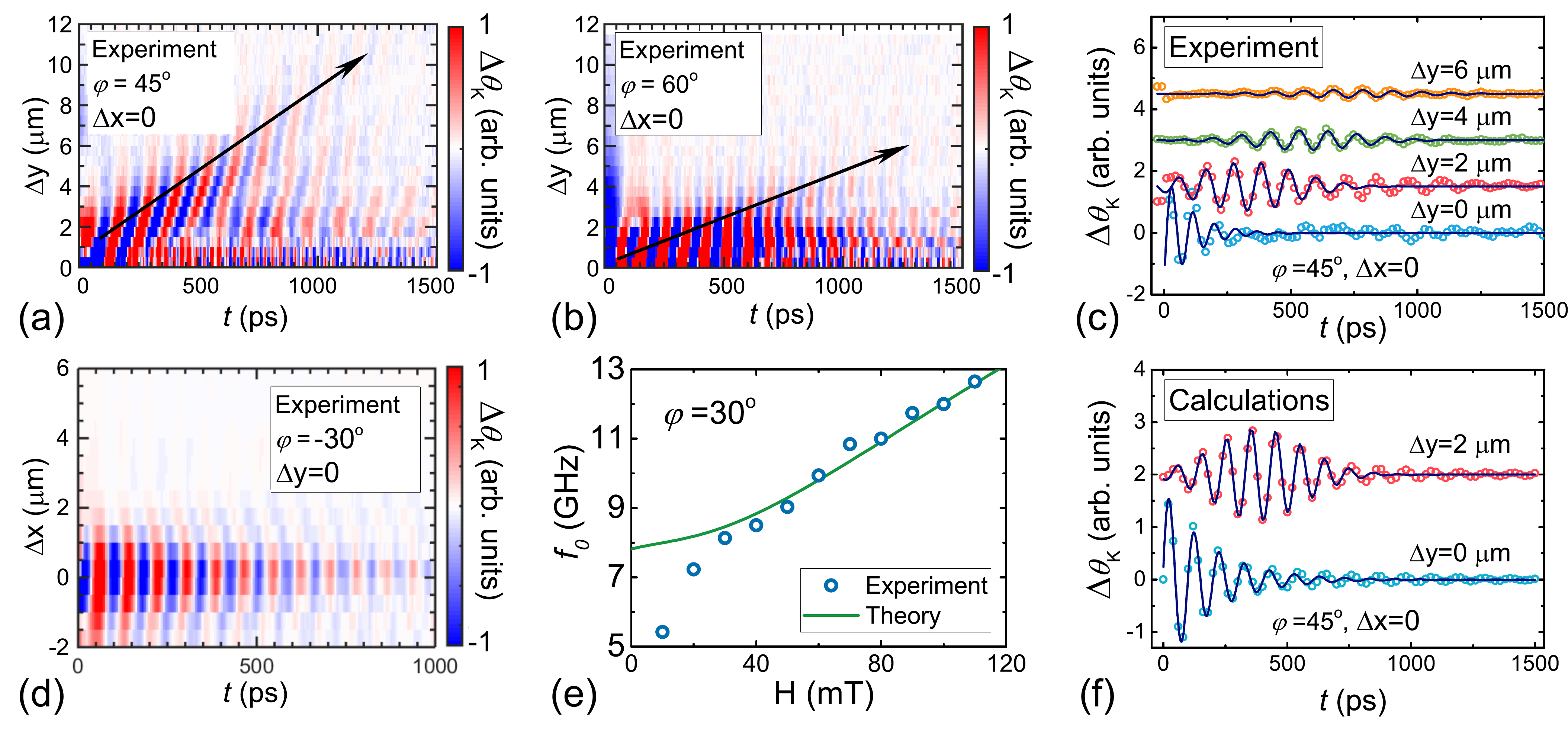}}
  \caption{ \label{fig2}
    (a,b) Experimental spatio-temporal $\Delta y-t$ maps of $\Delta\theta_K$ obtained at  $\varphi=45^{\circ}$ (a) and $\varphi=60^{\circ}$ (b) and $\Delta x=0$, i.e when  the  pump-probe distance $\Delta y$ is scanned transversely to {\bf H}, corresponding to the MSSW configuration.
  Arrows are guides to the eye showing the centers of the propagating MSSWs packets.
  (c) Experimental (symbols) temporal evolution of $\Delta\theta_K$ at different pump-probe distances $\Delta y$, at $\varphi=45^{\circ}$ and $\Delta x = 0$.
  Solid lines: the fits using Eqs.\,(\ref{eq:precession},\ref{eq:wavepacket}).
  (d) Experimental spatial-temporal $\Delta x-t$ map of $\Delta\theta_K$ obtained at  $\varphi=-30^{\circ}$ and $\Delta y = 0$, i.e. when the pump-probe distance $\Delta x$ is scanned along {\bf H}, corresponding to the BVMSWs configuration.
  (e) Experimental (symbols) and calculated using Eq.(\ref{eq:disper_simpl}) dependence of the precession frequency $f_0$ on the strength of the magnetic field $H$ applied at $\varphi=30^{\circ}$. 
  (f) Calculated (symbols) temporal evolution of $\Delta\theta_K$ at different pump-probe distances $\Delta y$, at $\varphi=45^{\circ}$ and $\Delta x = 0$. 
  Solid lines: the fits using Eqs.\,(\ref{eq:precession},\ref{eq:wavepacket}).
  }
\end{figure*}

\section{\label{sec:results}Experimental results and discussion}

\subsection{\label{sec:SWpackets}Laser-induced excitation of propagating spin waves}

Figures\,\ref{fig2}(a,b) show the spatial-temporal evolution of $\Delta\theta_K$ obtained by scanning the pump-probe time delay $t$ when the pump and probe spots are shifted with respect to each other by $0\leq\Delta y\leq12$\,$\mu$m at $\Delta x=0$, i.e. transversely to $\mathbf{H}$. 
Experimental data for two orientations of the sample $\varphi=45^{\circ}$ and $\varphi=60^\mathrm{o}$ are presented.
The former geometry corresponds to the field applied along the film hard axis [110]. 
Two types of pump-induced signal $\Delta\theta_K(t)$ can be distinguished depending on whether the pump and probe spots overlap spatially or not. 
We show this in more detail in Fig.\,\ref{fig2}(c), where the cross-sections of the spatial-temporal maps at various $\Delta y$ are presented.  
When the pump and probe spots overlap spatially, i.e. at $\Delta y< \sqrt2\sigma$, decaying oscillations of $\Delta\theta_K(t)$ are observed. 
Examination of the dependence of the oscillation frequency on the external field strength [Fig.\,\ref{fig2}(e)] confirms that they originate from the laser-induced precession of the magnetization. 
Outside the pump-probe spatial overlap, i.e. at $\Delta y>\sqrt2\sigma$, well-defined wave-packets are observed in the $\Delta\theta_K(t)$ signal.
The tilts of the signal maxima reveal a positive phase shift of the propagating waves. 
Therefore, these wave-packets can be confidently ascribed to laser-induced MSSWs propagating transversely to \textbf{H}.

Spatial-temporal $\Delta x-t$ maps obtained at $\Delta y=0$ (Fig.\,\ref{fig2}(d)) have revealed the presence of fast-decaying backward volume magnetostatic waves (BVMSWs) with negative phase shifts.
Such a difference in the propagation characters of laser-driven MSSWs and BVMSWs appears to be typical for thin metallic films \cite{YunAPEX:2015,IihamaPRB:2016,KamimakiPRB:2017}.
Below we focus our discussion on the excitation and propagation of the MSSWs.

As can be seen in Figs.\,\ref{fig2}(a,b), the parameters of the magnetization precession at $\Delta y=0$ and of the MSSWs at $\Delta y\neq0$ vary with $\varphi$. 
To quantify the azimuthal dependences of the parameters, we fitted temporal signals $\Delta \theta_K(t)$ obtained at different $\varphi$ and $\Delta y$ with either of the functions:
\begin{eqnarray}
\label{eq:precession}
  &\Delta \theta_K& (\Delta y=0, t) = A_\mathrm{SW}^0 
    \sin(2\pi f_0 t - \phi_0)e^{-t/\tau},\\
\label{eq:wavepacket}
&\Delta \theta_K &(\Delta y, t) = A_\mathrm{SW} (\Delta y) 
    \sin(2\pi f t - \phi)e^{-\frac{(t-t_0)^2}{2w^2}}.
\end{eqnarray}
Here $A_{\rm{SW}}^0$, $A_{\rm{SW}}(\Delta y)$ are amplitudes of the precession at $\Delta y=0$ and $\Delta y > 0$, respectively; 
$\tau$, $f_0$ and $f$,  $\phi_0$ and $\phi$ are the decay time, frequencies, and initial phases of the precession; $w$ and $t_0$ -- width and center position of the Gaussian packet.
Good agreement between the experimental data and the fitted curves is reached [Fig.~\ref{fig2}(c)], apart from a small discrepancy at the tails of the precession and wavepackets, which is addressed below in Sec.\,\ref{sec:theory}.

\subsection{\label{sec:parameters}Mechanism of MSSWs' excitation}

\begin{figure}
    \center{\includegraphics[width=1\linewidth]{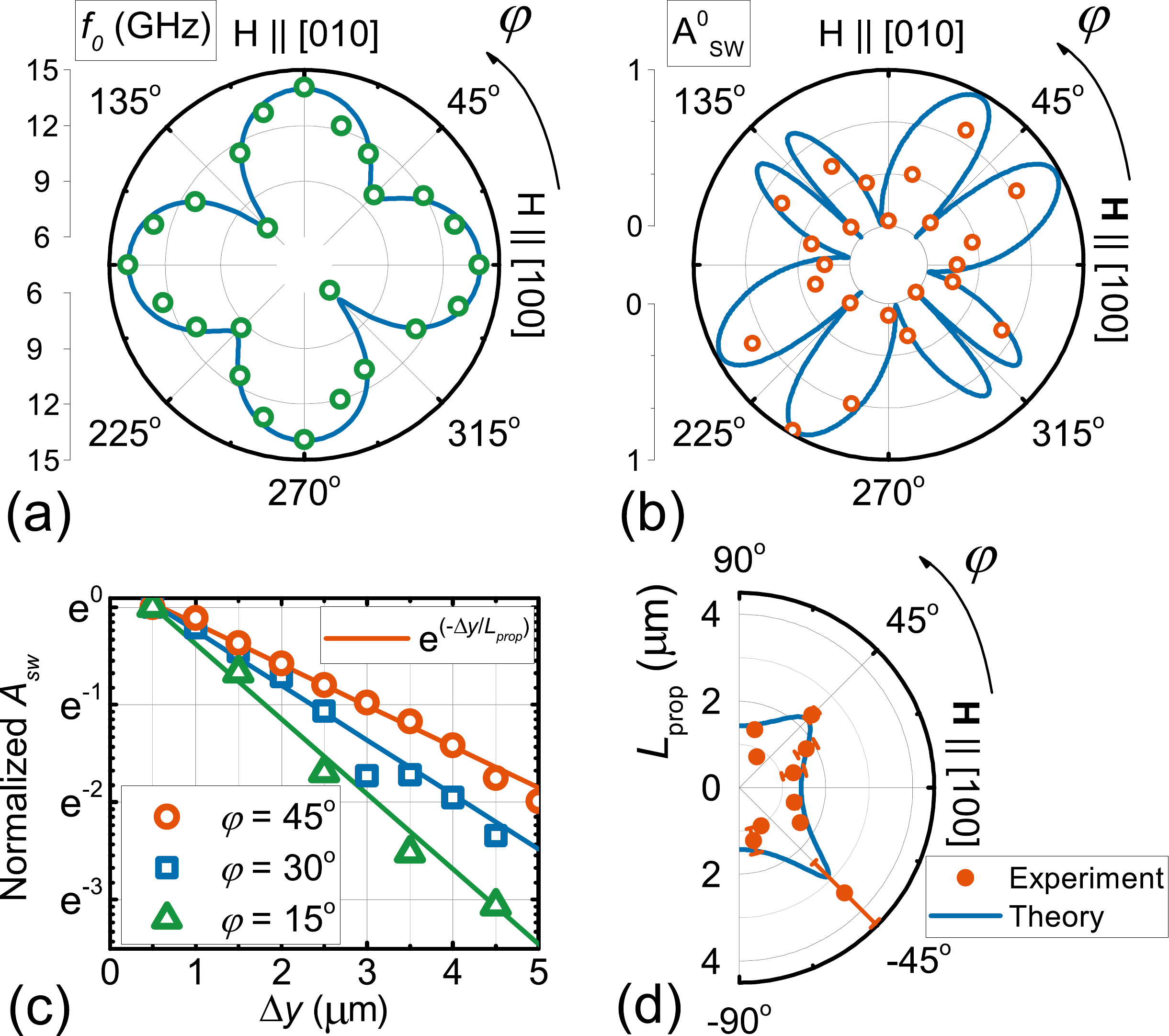}}
  \caption{ \label{fig3}
  (a,b) Azimuthal dependences of the frequency $f_0$ (a) and amplitude $A_{SW}^0$ (b) of the precession observed at $\Delta y=\Delta x=0$. 
  Symbols show experimental data; solid lines are the theoretical fits.
    (c) Normalized amplitude of MSSWs packets $A_\mathrm{SW}(\Delta y)/A_\mathrm{SW}(\Delta y=0.5\mu\mathrm{m})$ vs distance $\Delta y$ as obtained for several angles $\varphi$; lines are the fits using Eq.\,(\ref{eq:decay}).  
  (d) Propagation length $L_\mathrm{prop}$ of MSSWs packets vs $\varphi$ as obtained from the fits of the experimental (symbols) and using Eq. \ref{eq:lprop} (solid line).
  }
\end{figure}

Figures~\ref{fig3} (a,b) show the azimuthal dependences $f_0(\varphi)$ and $A_\mathrm{SW}^0(\varphi)$ at $\Delta y=0$. 
The former has a pronounced 4-fold symmetry with a 2-fold distortion.
It corresponds well to the intrinsic cubic magnetic anisotropy defined by the anisotropy parameter $K_1>0$ with easy axes along $[100]$ and $[010]$ and the substrate-induced in-plane uniaxial magnetic anisotropy defined by the parameter $K_u<0$ with an easy axis along $[110]$ expected for such films \cite{Wastlbauer:AP2005}.

A pronounced azimuthal dependence of the amplitude $A_\mathrm{SW}^0(\varphi)$ of the laser-driven precession [Fig. \ref{fig3}(b)] allows us to identify the excitation mechanism as an ultrafast laser-induced \textit{thermal}  change of the effective magnetocrystalline anisotropy field demonstrated earlier for a range of Galfenol films of various thicknesses \cite{KatsPRB:2016, Scherbakov:PhysRevAppl2019}. 
In brief, excitation of the precession stems from a rapid decrease of the anisotropy parameters $K_1$ and $K_u$ \cite{CarpenePRB:2010,MaJAP:2015,ShelukhinPRB:2018} and of the saturation magnetization $M_s$ \cite{Bigot:PRL1996} in response to the laser-induced increase of electronic and lattice temperatures. 
For the in-plane orientation of external magnetic field considered here, the amplitude of the excited precession is a measure of the abrupt reorientation of the total effective field in the sample plane due to the laser-induced changes of  $K_1/M_s$ and $K_u/M_s$.
At the timescale longer than the electron-phonon thermalization time $\sim$2\,ps, the thermodynamic approximation for the temperature dependence of anisotropy can be applied \cite{Zener-PRB1954}.
Thus, the decrease of the parameters $K_{1}$, $K_{u}$ is expected to be stronger than that of the magnetization $M_s$, and, therefore, the laser excitation yields an abrupt decrease of the effective anisotropy field.
Experimentally verified independence of the $\Delta\theta_k(t)$ signal on the pump pulse's polarization supports the thermal nature of the excitation mechanism.

In the following we refer to the MSSWs excitation mechanism based on laser-induced thermal change of the effective magnetocrystalline anisotropy field ${\sim(K_{1}+K_{u})/M_s}$ as the ultrafast laser-induced change of the magnetic anisotropy.
This is to distinguish it from the excitation of the MSSWs via ultrafast changes of the shape anisotropy related to the demagnetization solely and demonstrated in \cite{Au:2013, IihamaPRB:2016, vanKampen:PRL2002, YunAPEX:2015, KamimakiPRB:2017, ChenSciRep2017}.
The latter mechanism is deliberately excluded here by the chosen experimental geometry with an in-plane external magnetic field.
We can also exclude two other mechanisms of laser-induced MSSWs' excitation reported earlier.
Ultrafast inverse magneto-optical effects employed in \cite{JacklPRX:2017,Satoh:2012,Chernov:2017,Hashimoto:2017,JacklPRX:2017,SavochkinSciRep:2017,YoshimineStupakiewicz:JAP2014} are proven to be efficient in dielectric media mostly (for a review see e.g. \cite{KalashnikovaUFN2015}), and are pump-polarization sensitive which was not the case in our experiments.
Driving SWs by optically-excited elastic waves shown in \cite{Hashimoto:2017,HashimotoPRB:2018, OgawaPNAS:2015, HashimotoAPL:2018, Hashimoto:PhysRevApplied2019} can, in turn, work in any material with sufficiently strong magnetoelastic coupling at any laser-pulse polarization.
Furthermore, the magnetoelastic properties of Galfenol favor such a mechanism.
However, excitation of SWs by elastic waves becomes efficient only near the crossing of their dispersion curves \cite{Azovtsev:PhysRevB2018}, which is not realized in the studied film in the range of the applied magnetic fields used in the experiments (see details in Sec.\,\ref{sec:FFT}).
Nevertheless, in the experiments we have detected two acoustic waves seen as two concentric ripples independently propagating slower than the SWs, in analogy with the observations reported in \cite{Au:2013}.

\subsection{\label{sec:Lprop}MSSWs' propagation length}

The azimuthal dependence of the MSSW's amplitude $A_\mathrm{SW}(\varphi)$ at $\Delta y=0.5$ $\mu$m, i.e. close to the edge of the excitation spot, naturally resembles the one for $A_\mathrm{SW}^{0}$.
At $\Delta y > \sqrt{2}\sigma$ the situation changes drastically, as the amplitudes of the MSSWs packets outside the excitation spot are defined not only by the efficiency of excitation at the specific field direction [Fig.\,\ref{fig3}(b)], but by the spatial decay as well. 
To single out the latter contribution, we plot the spatial dependence of the normalized amplitudes of the MSSWs packets $A_\mathrm{SW}(\Delta y)/A_\mathrm{SW}(\Delta y=0.5$ $\mu$m) [Fig.\,\ref{fig3}(c)].  
The spatial decay is minimum at $\varphi = \pm45^{\circ}$ in the so-called "hard-hard" configuration, where the equilibrium magnetization and the MSSW's wavevector are oriented along the two orthogonal hard axes.
Thus, the MSSWs packets propagate larger distances along the hard axes, despite small initial amplitudes defined by the excitation mechanism.
In contrast, no propagating MSSWs packets are observed if {\bf H} is directed close to the easy axes ($\varphi = 0^{\circ}, \pm90^{\circ}$), and the small amplitude precession is excited.

The experimental dependences $A_\mathrm{SW}(\Delta y)$ can be well fitted by a single exponential decay function [Fig.\,\ref{fig3}(c)]:
 \begin{equation}  \label{eq:decay}
    A_\mathrm{SW}(\Delta y) \sim e^{-\Delta y/L_\mathrm{prop}},
 \end{equation}
with $L_\mathrm{prop}$ being the propagation length. 
This is an important parameter of SWs, which, in particular, determines whether Galfenol is a suitable material for magnonic applications.
As can be seen in Fig.\,\ref{fig3}(d), $L_\mathrm{prop}$ demonstrates a pronounced azimuthal dependence with two maxima corresponding to the geometries with {\bf H} applied along the hard axes.
The largest $L_\mathrm{prop}=3.4$\,$\mu$m is observed when {\bf H} is aligned along the hardest anisotropy axis $[1\overline{1}0]$.
Importantly, this value is very close to the propagation length found for optically excited MSSWs in a 20-nm thick Permalloy film \cite{IihamaPRB:2016}.

\subsection{\label{sec:FFT}Reconstruction of MSSW dispersion}

\begin{figure} 
    \center{\includegraphics[width=1\linewidth]{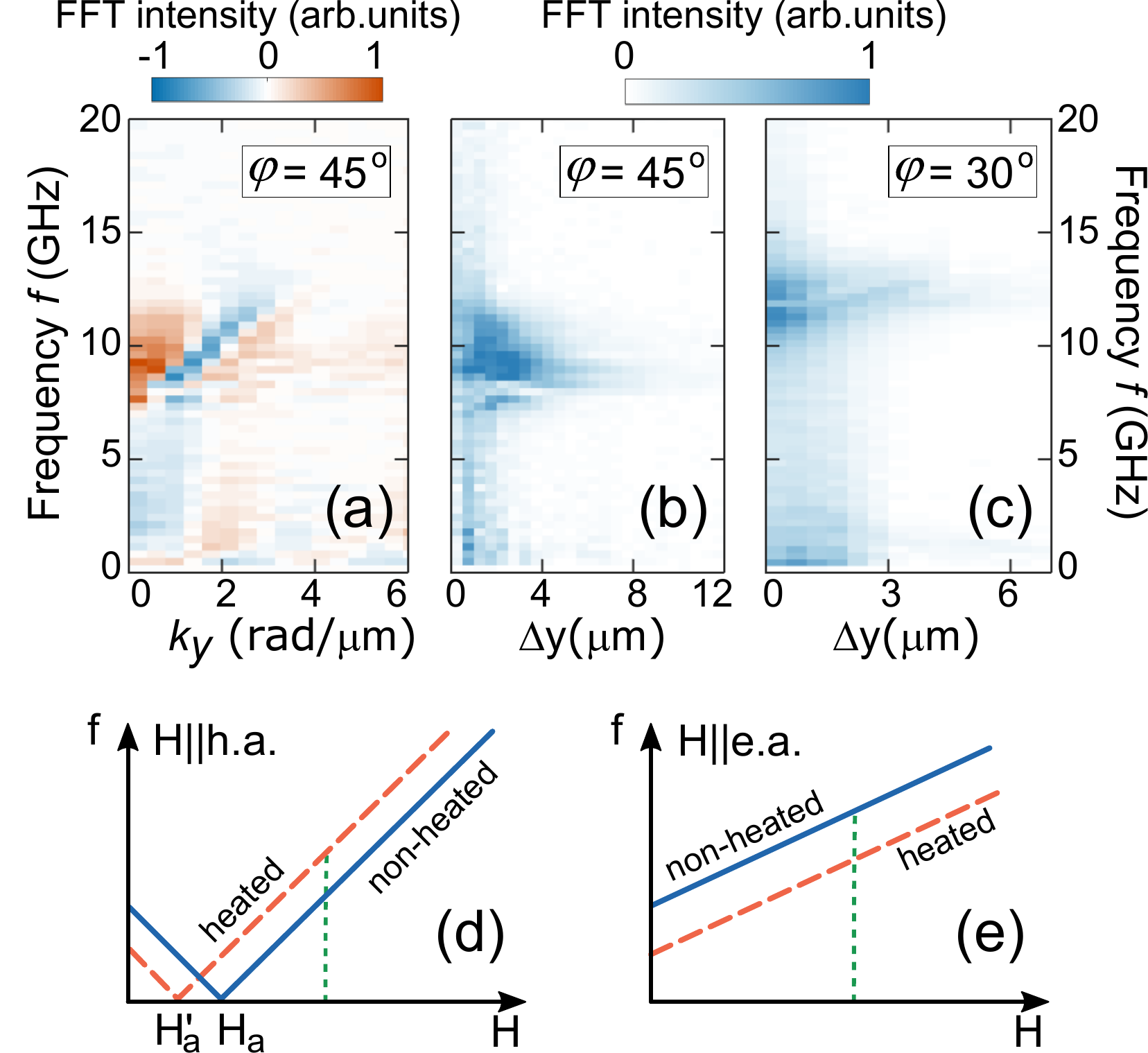}}
  \caption{ \label{fig:FFTmaps}
  (a) Dispersion of MSSWs reconstructed via 2D FFT (real part) at $\varphi=45^{\circ}$.
  (b,c) FFT spectra at different pump-probe shifts $\Delta y$ at $\varphi=45^{\circ}$ and $30^{\circ}$.
  (d,e) Schematic representation of the frequency changes in the heated area when {\bf H} is along a hard and easy anisotropy axis, respectively.
  Solid lines: dependences of $f(H)$ for a non-heated film, dashed lines: $f'(H)$ for a heated film. 
  Vertical dashed lines represent the magnitude of {\bf H} in the experiments.}
\end{figure}

The $\Delta y-t$ maps presented in Figs.\,\ref{fig2}(a,b) allow us to reconstruct the dispersion relation $f(k_y)$ for the excited MSSWs using a 2-dimensional fast Fourier transform (2D-FFT) of the data outside the pump spot \cite{KamimakiPRB:2017,Hashimoto:2017,HashimotoAPL:2018}.
Figure\,\ref{fig:FFTmaps}(a) shows the {2D-FFT} result for $\varphi=45^{\circ}$, at which the MSSWs possess large $L_{\rm prop}$ (see Sec.\,\ref{sec:Lprop}). 
The 2D-FFT gives an expected near-linear dispersion and shows that the wavenumbers $k_y$ of the MSSWs detected in the experiment reach about 3.5 rad/$\mu$m.
As discussed in Ref.\,\cite{KamimakiPRB:2017} the wavenumbers of the SWs detected in the optical pump-probe experiment are limited by  both the pump and probe spot sizes, which yields $k_{\sigma}=1/\sigma$ for the wavenumbers at the $1/\sqrt{e}$-level (see Appendix for details). 
Given the spot size $\sigma$ used in the experiment, the corresponding value is $k_{\sigma} = 1.3\,$\,rad/$\mu$m, and the experimentally found largest value 3.5 rad/$\mu$m corresponds approximately to $3k_{\sigma}$.
The value also confirms that the observed MSSWs are not driven by propagating surface acoustic waves (SAWs) in the GaAs substrate.
Indeed, the SAWs velocity in a (001)-GaAs substrate is 2.7\,km/s \cite{Maznev:EPJB2003}, which gives a maximum SAW frequency of 1.5 GHz at a wavenumber of 3.5\,rad/$\mu$m.
The value is well below the frequencies of the excited MSSWs.

It is important to note that the laser-induced heating and the changes of magnetization and anisotropy are expected to alter the precession frequency $f$ within the pump spot.
These changes, indeed, can be clearly seen in $f - \Delta y$ maps, where $f$ is obtained as the FFT of $\theta_{\rm K}(t)$ at a specific position $\Delta y$ [Fig.\,\ref{fig:FFTmaps}(b,c)].
For {\bf H} applied along the hard axis ($\varphi=45^{\circ}$) $f$ experiences an increase inside the heated area. 
If {\bf H} is not aligned with the hard axis (e.g. at $\varphi=30^{\circ}$), the situation is opposite, and $f$ is decreased inside the pump spot.
Both observations are in agreement with the scenario of the magnetic anisotropy being decreased abruptly within the laser-excited spot. 
Indeed, in the "hard-hard" configuration the laser-induced ultrafast heating partly suppresses the effective anisotropy field $H_a$ resulting in an increase of $f$ at $H>H_a$, as schematically illustrated in Fig.\ref{fig:FFTmaps}(d).
The opposite situation is expected in the "easy-easy" configuration [see Fig.\ref{fig:FFTmaps}(e)]. 

It should be noted that the laser-induced local decrease of the eigen frequency $f_0$ can lead to the formation of a potential well for MSSWs, with eigen frequencies below the spectrum of MSSWs outside this well, as demonstrated in Ref.\,\cite{Kolokoltsev:APL2012}.
Then the escape of the MSSWs from this well would be strongly suppressed.
We do not observe the formation of such traps as they are formed, when the diameter of the hot spot is several times larger than the MSSW wavelength.
In our pump-probe experiments the shortest excited MSSW wavelength is comparable to the pump spot size.
In other words, when the pump spot size becomes comparable to the length scale defined by the ratio of the group velocity of a MSSW to $f_0$, the precession of the magnetization propagates as MSSWs from the pump spot to the non-excited area \cite{KamimakiPRB:2017}.

\subsection{\label{sec:theory}Theoretical analysis}
\subsubsection{Dispersion of magnetostatic waves in an anisotropic film}

In order to obtain the expression for the spin wave dispersion $f(\mathbf{k})$ we utilized the approach developed in Ref.\,\cite{HurbenJMMM1996} and adapted it to the particular anisotropy of the studied film.
The basic expression for the film's free energy density used in our calculations has the form:
\begin{multline}\label{eq:free_energy}
  \Delta F = K_1(m_1^2 m_2^2 + m_1^2 m_3^2 + m_2^2 m_3^2)\\
  + K_u m_1 m_2 - \mu_0 M_s\textbf{m}\cdot\textbf{H},
\end{multline}
where $\textbf{m}$ is the unit vector in the magnetization direction, and $m_i$ is its projection on the crystallographic axis $x_i$ [see inset in Fig.\,\ref{fig1}].
The form of the in-plane anisotropy term $K_um_1m_2$ was chosen after Ref. \cite{Wastlbauer:AP2005} to accommodate our system of an iron-based alloy grown on a (001)-GaAs substrate.
Note that Eq. (\ref{eq:free_energy}) does not contain the magnetostatic energy term.
In the derivation of the dispersion relation the demagnetizing field is accounted for separately by solving the Maxwell's equations. 
Applying the procedures of Ref.\,\cite{HurbenJMMM1996} to the free enery density given by Eq.\,\ref{eq:free_energy}, we obtained a transcendental equation connecting $f$ and $\textbf{k}$, which can be simplified and written in a closed form in the limit $kd \ll 1$ relevant in our case \cite{Walker:1957}, yielding the dispersion relation for magnetostatic waves to be:

\begin{multline} \label{eq:disper_simpl}
\omega(\mathbf{k}) = 2\pi f(\mathbf{k}) =\\= \gamma \sqrt{\Big(B_{\alpha} + \mu_0M_s(1 - \frac{kd}{2})\Big)\Big(B_{\beta} + \mu_0M_s\frac{kd}{2}\sin^2{\psi}\Big)},
\end{multline}
with $B_{\alpha}$ and $B_{\beta}$ defined as
\begin{multline*} \label{eq:HaHb}
B_{\alpha} = \mu_0 H\cos(\varphi - \varphi_m) - \frac{K_u}{M_s}\sin(2\varphi_m) \\
+ \frac{2K_1}{M_s}\left[1 - \frac{1}{2}\sin^2(2\varphi_m)\right];
\end{multline*}
\begin{multline}
B_{\beta} = \mu_0 H\cos(\varphi - \varphi_m) - \frac{2K_u}{M_s}\sin(2\varphi_m) \\
+ \frac{2K_1}{M_s}\cos(4\varphi_m),
\end{multline}
where $\psi$ is the angle between the equilibrium  magnetization and $\mathbf{k}$ [see inset in Fig.\,\ref{fig1}], $\gamma=1.76\times10^{11}$\,rad$\cdot$s$^{-1}\cdot$T$^{-1}$ is the electron's gyromagnetic ratio, and $\varphi_m$ is the angle between the magnetization and the [100] axis obtained from the equilibrium condition:
\begin{multline} \label{eq:equilibrMS}
 \mu_0 H\sin(\varphi - \varphi_m) - \frac{K_1}{2M_s}\sin(4\varphi_m) \\- \frac{K_u}{M_s}\cos(2\varphi_m) = 0.
\end{multline}

The discrepancy between the solution of the exact transcendental equation for $f(\mathbf{k})$ and its approximation (\ref{eq:disper_simpl}) is negligible at $k < 5$ rad/$\mu$m and $d = 20$ nm. 
We note that at $k=0$, i.e. in the case of a homogeneous magnetization precession, Eq. (\ref{eq:disper_simpl}) gives exactly the same dependence $f_0(\varphi)$ as the analytical formula for ferromagnetic resonance frequency in a thin anisotropic film \cite{AzovtsevAV:PRA2018}.
\par Having obtained the explicit dispersion relation, we derived the expression for the precession ellipticity:
\begin{equation} \label{eq:ellipticity}
\epsilon = \frac{|\Delta M_{xy}|}{|M_z|} = \sqrt{\frac{|B_{\alpha} + \mu_0M_s(1 - kd/2)|}{|B_{\beta} + \mu_0M_s(kd/2)\sin^2{\psi}|}},
\end{equation}
where $\Delta M_{xy}$ and $M_z$ are the transient changes of in-plane and out-of-plane components of the magnetization precession, respectively.
The damping factor as a function of $\varphi$ and $\bf{k}$ can be found using the relation Ref.\,\cite{Stancil:2009}:
\begin{equation} \label{eq:damping}
\alpha= \alpha_0 \frac{1}{\mu_0 \gamma} \frac{\partial \omega({\bf k})}{\partial H},
\end{equation}
where $\alpha_0$ is the Gilbert damping of the precession in the external field {\bf H} at $k=0$  without the account of demagnetization effects and anisotropy.
It is interesting to note that Eq.\,(\ref{eq:damping}) leads to an anisotropic damping with higher and lower values when the magnetization is aligned along hard and easy axes, respectively.

Once the Gilbert damping parameter is known, the formula for the propagation length $L_{\text{prop}}$ can be written in a straightforward way as:
\begin{equation} \label{eq:lprop}
L_{\text{prop}} = \tau v_{\text{gr}} = \frac{1}{\alpha \omega(\textbf{k})} \frac{\partial \omega(\textbf{k})}{\partial \textbf{k}},
\end{equation}
where $\tau$ is the relaxation time and $v_{\text{gr}}$ is the group velocity of the MSSW.

\subsubsection{Analysis of the magnetization precession within the excitation spot}

In order to apply the expressions derived above for a description of the experimentally observed magnetization dynamics, one needs to account for the timescales of the laser-induced changes of magnetic anisotropy and magnetization. 
In metals a few picoseconds after excitation both magnetization and magnetic anisotropy possess slow relaxations which can be neglected on the timescales where the precession is observed.
Thus, the temporal profile of the total effective field ${\sim(K_{1}+K_{u})/M_s}$ is described below by the Heaviside function. 

By analyzing the experimental dependences $f_0(\varphi)$ and $A_{SW}^0(\varphi)$ within the pump spot, we extract the anisotropy parameters $K_{1},K_u$ and saturation magnetization $M_s$ of the excited and equilibrium film, using the following procedure.
First, the frequency $f_0$ within pump spot is defined by modified parameters $K_1$, $K_u$, and $M_s$ of the laser-exited material. 
Fitting the experimental dependence $f_0(\varphi)$ to  Eq.\,(\ref{eq:disper_simpl}) with $k=0$ [solid line in Fig.\ref{fig3}(a)] we get $\mu_0 M_s=1.56$\,T, $K_1 = 2.8\times10^4$\,J/m$^3$ and $K_u = -1\times10^4$\,J/m$^3$.

Having determined the parameters of the film within the laser-excited area, we now can extract the laser-induced changes of the anisotropy parameters $K_1$ and $K_u$ with respect to their room temperature (RT) values by analyzing the azimuthal dependence $A_{\rm SW}^0(\varphi)$ of the laser-excited precession [Fig.\ref{fig3}(b)].
Indeed, two sets of the parameters, the initial $\tilde{K_1}$, $\tilde{K_u}$ and $\tilde{M_s}$ at RT and the modified $K_1$, $K_u$ and $M_s$ of the laser-excited film, yield the difference between the effective field orientation at equilibrium and upon the excitation.  
This difference defines the amplitude of the in-plane component $\Delta M^0_{xy}$ of the precession. 
The amplitude of the out-of-plane component $\Delta M^0_z$ can be then found taking into account the precession ellipticity (\ref{eq:ellipticity}).
The fitting of $A_{\rm SW}^0(\varphi)\sim\Delta M^0_z$ then gives
the ratios $\tilde{K_1}/\tilde{M_s}$ and $\tilde{K_u}/\tilde{M_s}$.
Using the experimental value of the RT saturation magnetization $\mu_0\tilde{M_s} = 1.7$\,T \cite{Gopman:IEEE2017, Parkes:SciRep2013}, we obtain the RT anisotropy parameters $\tilde{K_1} = 3.33\times10^4$\,J/m$^3$ and $\tilde{K_u} = -1.03\times10^4$\,J/m$^3$, which are in a very good agreement with the anisotropy parameters found earlier for a 22-nm FeGa film on a (001) GaAs substrate \cite{Parkes:SciRep2013}. 
We note that the laser-induced change $\Delta K_1/\tilde{K_1}\approx-16\%$ shows rather good agreement with the one reported earlier for a 100-nm Galfenol film \cite{KatsPRB:2016}, while $\Delta M_s/\tilde{M_s}\approx-8\%$ agrees with the magnitude of the ultrafast demagnetization observed in a thinner Galfenol film \cite{Scherbakov:PhysRevAppl2019}.

\subsubsection{\label{sec:theoryProp}Analysis of the laser-driven spin waves propagation}

In order to explain the observed azimuthal dependence $L_\mathrm{prop}(\varphi)$, we used Eq.\,(\ref{eq:lprop}).
The fitting procedure at $k=1$rad/$\mu$m gives the Gilbert damping parameter $\alpha_0 = 0.017$.
As can be seen in Fig.\,\ref{fig3}(d),  Eqs.\,(\ref{eq:damping},\ref{eq:lprop}) correctly predict the azimuthal dependence of MSSWs' propagation.
In particular, the theory confirms that the different propagation lengths for different misorientations between the \textbf{H} and MSSW's wavevector are dictated by the anisotropy of the film.
We note that the observed anisotropy of the propagation of the laser-driven MSSWs packets is in agreement with the one demonstrated in cubic iron films recently \cite{SekiguchiNPGAsia:2017}. 
In both cases the largest propagation length is observed in the "hard-hard" configuration.
In contrast to the experiments with antennae \cite{SekiguchiNPGAsia:2017}, where the MSSWs' propagation is one-dimensional, laser-induced MSSWs packets propagating along the $y-$axis are spreading slightly along the $x-$axis as well.
Our results show that this spreading does not change the main features of the propagation of MSSWs packets.

Theoretically obtained MSSWs' group velocities $v_{\text{gr}}$ are in the range of 5-9\,km/s, and are in an agreement with the experimental values of 4.5-13 km/s.
These values of group velocities are typical for MSSWs in thin metallic films \cite{ChumakJPD2017}.

Finally, having obtained the MSSWs' parameters, including the anisotropic damping $\alpha$, we have calculated spatial-temporal $\Delta y-t$ maps at different $\varphi$, following the procedure described in \cite{Satoh:2012,SavochkinSciRep:2017,Hashimoto:2017,KamimakiPRB:2017}. 
In this approach the dispersion relation (\ref{eq:disper_simpl}) for MSSWs was used.
Within the area corresponding to the excitation spot the additional effective field with the Heaviside temporal and Gaussian spatial profiles were introduced to account for the localized change of magnetic anisotropy triggering the precession (see Appendix for details).
Figure\,\ref{fig2}(f) shows exemplary crossections of the calculated $\Delta y-t$ maps demonstrating good agreement with the experimental data.
The calculations show good agreement with experimentally obtained precession and MSSWs packets [Fig.\,\ref{fig2}(c,f)].
Importantly, the calculations reveal the same deviation of the MSSW packets' shapes from the Gaussian ones as the experimentally observed signals $\Delta \theta_{\rm K}(t)$, evident in Figs.\,\ref{fig2}(c,f).
The most prominent deviation is observed at large time delays $t$, i.e. at the tails of the wavepackets, as was noted by other groups as well \cite{MingzhongJAP:2006, IihamaPRB:2016, KamimakiPRB:2017}. 
Our calculations based on the MSSWs' dispersion relation show that this deviation originates from the dispersion of SWs generated by a sudden change of effective field.

\section{\label{sec:Conclusion}Conclusion}

\begin{figure*}
    \center{\includegraphics[width=1\linewidth]{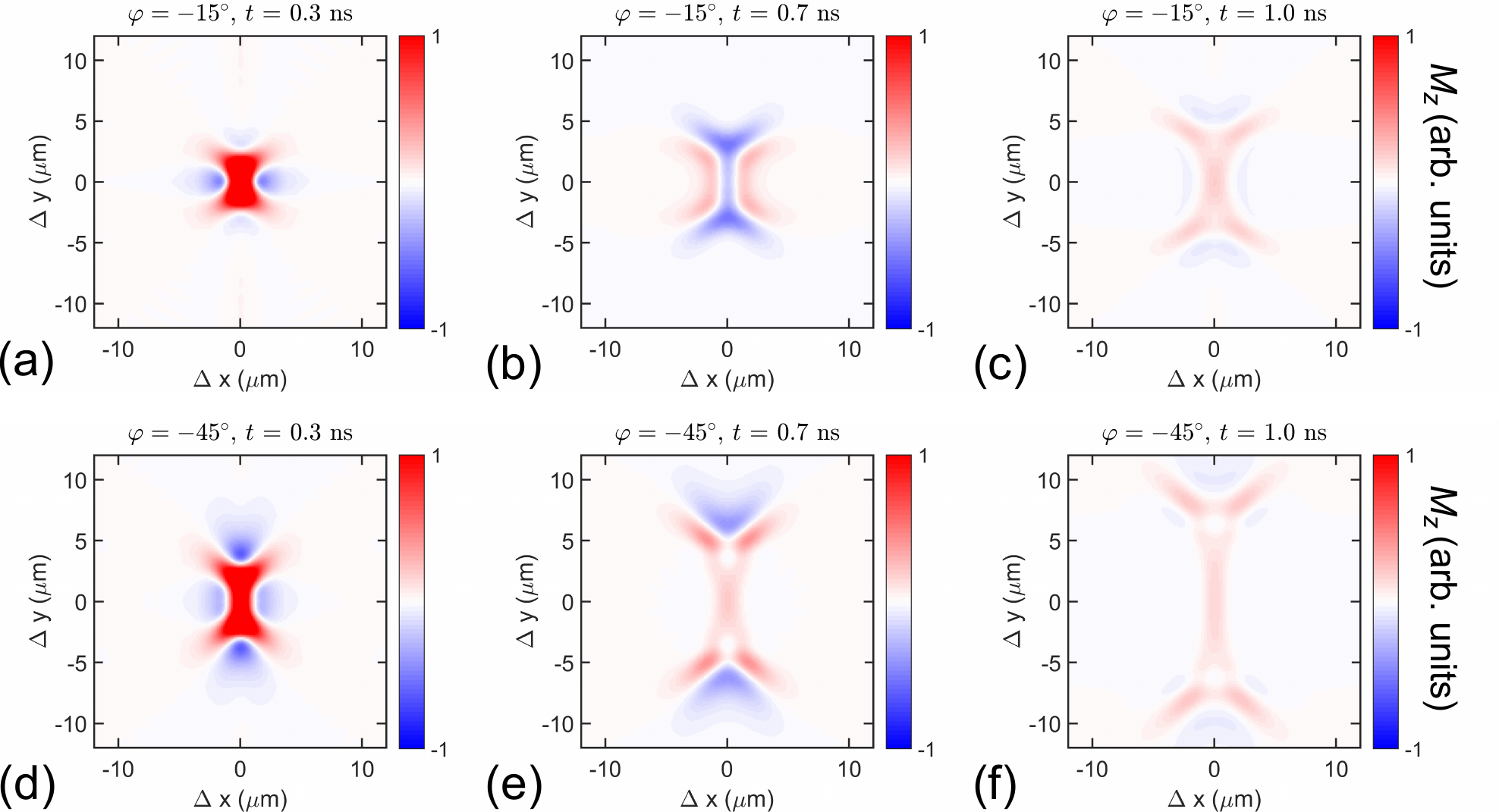}}
\caption{\label{XYmaps}
Calculated SWs signals in the $xy$ plane at different time delays $t$ after the excitation, 0.3 ns (a,d), 0.7 ns (b,e), and 1 ns (c,f) for the angles $\varphi$ of -15$^{\circ}$ (a-c) and -45$^{\circ}$ (d-f).
}
\end{figure*}

In conclusion, we have demonstrated laser-induced excitation and propagation of magnetostatic spin waves in a 20~nm thick epitaxial Galfenol film on a GaAs substrate characterized by pronounced in-plane magnetic anisotropy.
We have shown that ultrafast thermal magnetic anisotropy changes induced by tightly focused femtosecond laser pulses excite propagating MSSWs.
The strong in-plane anisotropy ($\sim$10$^4$\,J/m$^3$) of the film enables the laser-induced excitation of MSSWs in a simple geometry with an in-plane external magnetic field.
The anisotropy of the film provides the possibility to tune the frequency, amplitude and propagation length of the excited waves by changing the in-plane field orientation.
We find that the propagation length of the MSSWs in the studied film reaches 3.4\,$\mu$m, which, along with other recent results on spin dynamics in Galfenol films \cite{DanilovPRB:2018,SalasyukPRB:2018, Scherbakov:PhysRevAppl2019}, confidently promotes epitaxial Galfenol to the limited family of metallic materials for magnonics. 
Furthermore, epitaxial Galfenol is also known for having a large magnetostriction constant \cite{Parkes:SciRep2013} and so offers the prospect of controlling the SWs propagation via a voltage-induced strain when, for example, coupled to a piezoelectric substrate, as was earlier realized in YIG \cite{Sadovnikov:PRL2018}.
It is important to note that the laser-induced thermal magnetocrystalline anisotropy change represents a novel fundamental  process for the excitation of SWs, which can be applied to a broad range of materials without limitations on their electronic and magnetic structures.
Finally, we note that introducing laser-induced ultrafast thermal changes of magnetic anisotropy as a tool to generate SWs can have even broader impact on magnonics.
Since abrupt and local changes of the magnetic anisotropy by laser pulses yield strong local modifications of the SWs' dispersion relation, they can be seen as a pathway to realize an ultrafast optically-reconfigurable magnonic medium for efficient steering and conversion of SWs \cite{StigloherPRL:2016,VogelSciRep:2018,VogelNPhys:2015, Sadovnikov:PRB2019}.

\section*{\label{sec:Acknowledgments}Acknowledgments}
The authors thank L. V. Lutsev and I. V. Savochkin for valuable discussions.
The setup construction and experiments were carried out under a support of RSF (grant No.\,16-12-10485); theoretical  analysis and calculations were performed under a support of RFBR (grant No.\,18-02-00824). Collaboration between Ioffe Institute and TU Dortmund is a part of the TRR 160 ICRC "Coherent manipulation of interacting spin excitations in tailored semiconductors" supported jointly by RFBR (grant No. 19-52-12065) and DFG, and of the COST action CA17123 Magnetofon. 

\section*{\label{Append:dispertion}Appendix: Calculation of spatial-temporal maps\\ of laser-driven spin waves}

To calculate the spatial-temporal dependences of $\Delta\theta_K (\mathbf{r},t)$ corresponding to the experimental $\Delta y-t$ maps,  we used the procedure described in Refs.\,\cite{Satoh:2012,SavochkinSciRep:2017,Hashimoto:2017,KamimakiPRB:2017}.
The probe pulses are reflected normally, so the change of their polarization is a measure of the transient changes of the out-of-plane component $M_z$.
The normal component is given by an integration over all excited wavevectors $\mathbf{k} = (k_x, k_y)$ \cite{KamimakiPRB:2017}:
\begin{multline}
    M_z(\mathbf{r},t) \sim \int d\mathbf{k} (1/\epsilon) h(\mathbf{k}, \omega) \sin [\mathbf{k}\mathbf{r} - \omega(\mathbf{k})t] \\ \times \exp[-\alpha \omega(\mathbf{k})t],
\end{multline}
where $h(\mathbf{k},\omega)$ is the Fourier transform of $h(\mathbf{r},t)$ - an effective field describing the effect of the laser pulse excitation of the sample, $\epsilon$ is the ellipticity (\ref{eq:ellipticity}), $\alpha$ is the  damping parameter (\ref{eq:damping}).

We consider the case of \textit{thermal} changes of anisotropy, which are approximated by the laser-induced effective field introduced as the difference between the total effective field in the equilibrium and laser-excited states
$h(\mathbf{r},t)=-\partial\Delta F(\tilde{K_1},\tilde{K_u},\tilde{M_s})/\partial\mathbf{M}+\partial\Delta F(K_1(\mathbf{r},t),K_u(\mathbf{r},t),M_s(\mathbf{r},t))/\partial\mathbf{M}$ \cite{ShelukhinPRB:2018}.
It is assumed that a few picoseconds after excitation both the magnetization and magnetic anisotropy possess slow relaxation which can be neglected on the time scales of the magnetization precession.
Thus, the field $h(\mathbf{r},t)$ may be rewritten as $h(\mathbf{r},t) \sim h(\mathbf{r})\Theta(t)$, where $\Theta(t)$ is a Heaviside function.
Hence, the Fourier transform $h(\mathbf{k},\omega)$ at $\omega \neq 0$ is weighted by $i/ \omega$ \cite{Hashimoto:2017}.
Next, the pump beam in our experiments has a spatial profile close to the Gaussian one, i.e. $h(\mathbf{r}) \sim \exp (-\mathbf{r}^2 / 2\sigma^2)$. 
Therefore, $h(\mathbf{k},\omega) \sim \exp(-\mathbf{k}^2\sigma^2/2)$.
Thus, $k'_{\sigma}=\sqrt2/\sigma$ gives a wavenumber for the excited SW component which amplitude is at a level $1/\sqrt{e}$ of the component with the highest amplitude.
It is also important to take into account that the probe spot radius in our experiment is $\sigma$, as well.
As a result, the measured wavevnumbers are defined by an effective width of $\sqrt2\sigma$ \cite{KamimakiPRB:2017}, and the corresponding wavenumber is $k_{\sigma}=1/\sigma$. 

Using the procedure described above, the maps of $M_z(\mathbf{r},t)$ were calculated in the time interval 0--1500\,ps for different $\varphi$.
Calculations were performed for $k_x, k_y$ with an upper limit of 5.3 rad/$\mu$m which exceeds the value $3k_{\sigma}$.
Calculated $\Delta y-t$ maps are in a good agreement with the experiment.

Figure\,\ref{XYmaps} shows the calculated $z$-component of the magnetization as a function of $\Delta x, \Delta y$ at different time delays $t=$0.3, 0.7, and 1\,ns (see also supplementary animation file).
Excited SWs are mainly propagating in the {$y$-direction} with a pronounced X-shaped pattern.
Similar {2-dimensional} maps for a Permalloy film are discussed in the Supplemental Material of Ref.\,\cite{IihamaPRB:2016}.
Slight spreading of the SWs in the {$x$-direction} is clearly seen.

%

\end{document}